\documentclass[graphicx,superscriptaddress,prl,reprint,longbibliography]{revtex4-1}

\usepackage[version=3]{mhchem} 
\usepackage[T1]{fontenc}
\usepackage{siunitx}
\usepackage{graphicx}
\usepackage[utf8]{inputenc}

\usepackage[normalem]{ulem}
\useunder{\uline}{\ul}{}
\begin{document}
\author{Robin J. Dolleman}
\email{R.J.Dolleman@tudelft.nl}
\altaffiliation[Current affiliation:]{ 2nd Institute of Physics, RWTH Aachen University, 52074 Aachen, Germany}
\affiliation{Kavli Institute of Nanoscience, Delft University of Technology, Lorentzweg 1, 2628 CJ, Delft, The Netherlands}
\author{Gerard J. Verbiest}
\affiliation{Department of Precision and Microsystems Engineering, Delft University of Technology, Mekelweg 2, 2628 CD, Delft, The Netherlands}
\author{Yaroslav M. Blanter}
\affiliation{Kavli Institute of Nanoscience, Delft University of Technology, Lorentzweg 1, 2628 CJ, Delft, The Netherlands}
\author{Herre S. J. van der Zant}
\affiliation{Kavli Institute of Nanoscience, Delft University of Technology, Lorentzweg 1, 2628 CJ, Delft, The Netherlands}
\author{Peter G. Steeneken}
\email{P.G.Steeneken@tudelft.nl}
\affiliation{Kavli Institute of Nanoscience, Delft University of Technology, Lorentzweg 1, 2628 CJ, Delft, The Netherlands}
\affiliation{Department of Precision and Microsystems Engineering, Delft University of Technology, Mekelweg 2, 2628 CD, Delft, The Netherlands}

\title{Nonequilibrium Thermodynamics of Acoustic Phonons in Suspended Graphene}

\begin{abstract}
Recent theory has predicted large temperature differences between the in-plane (LA and TA) and out-of-plane (ZA) acoustic phonon baths in locally-heated suspended graphene. To verify these predictions, and their implications for understanding the nonequilibrium thermodynamics of 2D materials, experimental techniques are needed.
Here, we present a method to determine the acoustic phonon bath temperatures from the frequency-dependent mechanical response of suspended graphene to a power modulated laser. The mechanical motion reveals two counteracting contributions to the thermal expansion force, that are attributed to fast positive thermal expansion by the in-plane phonons and slower negative thermal expansion by the out-of-plane phonons.  The magnitude of the two forces reveals that the in-plane and flexural acoustic phonons are at very different temperatures in the steady-state, with typically observed values of the ratio $\Delta T_{\mathrm{LA+TA}}/\Delta T_{\mathrm{ZA}}$ between 0.2 and 3.7. These deviations from the generally used local thermal equilibrium assumption ($\Delta T_{\mathrm{LA+TA}}=\Delta T_{\mathrm{ZA}}$) can affect the experimental analysis of thermal properties of 2D materials.
\end{abstract}
\maketitle

The thermal properties of graphene \cite{geim2010rise} are unconventional, because of the large difference between its in-plane and out-of-plane lattice dynamics \cite{nika2012two,pop2012thermal}. 
Therefore, much research has focused on characterizing graphene's thermal conductivity, for example by using Raman spectroscopy or electrical heaters \cite{balandin2008superior,chen2012thermalb,cai2010thermal,ghosh2010dimensional,seol2010two,lee2011thermal,xu2014length,faugeras2010thermal}.
Recent theoretical work by \citeauthor{vallabhaneni2016reliability} has suggested that local optical heating of suspended graphene can lead to a large temperature differences between the in-plane (longitudinal LA and transverse TA) and out-of-plane (flexural, ZA) acoustic phonon baths, which is caused by differences in the thermal conductivities of the different types of phonons, and their weak mutual interactions \cite{vallabhaneni2016reliability}. It has been confirmed experimentally that electrons and optical phonons can show very different temperatures compared to the acoustic phonons in 2D materials \cite{wang2010ultrafast,kim2015bright,chae2009hot,berciaud2010electron,block2019tracking,sullivan2017optical}, but whether strong thermal nonequilibrium between the acoustic phonon modes themselves exists has not been established. Since it has been hypothesized that such a thermal nonequilibrium might impact the interpretation of the widely used Raman spectroscopy technique to measure the thermal conductivity of graphene \cite{vallabhaneni2016reliability}, there is a need to characterize the temperatures of the in-plane and flexural acoustic phonon baths separately.
  
Recently, several optomechanical techniques to characterize the time-dependent heat transport in suspended 2D materials have been developed \cite{PhysRevB.96.165421,dolleman2018transient,morell,blaikie2018fast}.
Here, we demonstrate the use of an optomechanical technique to distinguish two thermal expansion force contributions with different time-constants and opposite signs. It is argued that these contributions can be attributed to the in-plane and flexural acoustic phonons. The differences in time-constant and sign allow us to obtain information on the modal temperatures of the respective phonon baths. 

\begin{figure}
\centering
\includegraphics{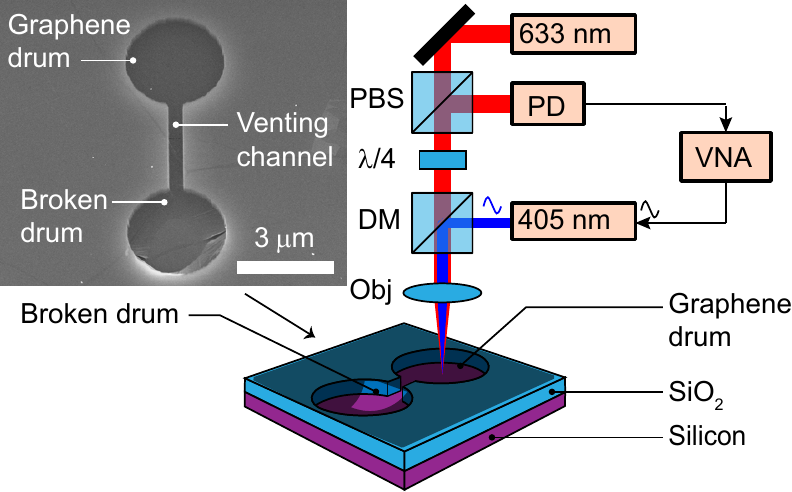}
\caption{Scanning electron microscope image of a typical device used in the experiment and the experimental setup to measure the thermomechanical response of suspended graphene membranes. \label{fig:gpmsetup}} 
\end{figure}
Figure \ref{fig:gpmsetup} shows the device and setup that is used to thermally actuate and measure the motion of suspended graphene membranes. The sample fabrication is identical to that in earlier work \cite{PhysRevB.96.165421}. Single-layer graphene grown by chemical vapor deposition (CVD) is transferred over dumbbell-shaped cavities in a Si/SiO$_2$ substrate (300 nm deep, various diameters) using a support polymer. The polymer is dissolved and the sample is dried using critical point drying, which breaks one of the dumbbell drums while the other side survives, resulting in a circular graphene drum with a venting channel to the environment that prevents gas from being trapped between the membrane and the substrate. 

To actuate the motion, the membrane is heated in a vacuum (pressure lower than $\num{1e-5}$ mbar) by a sinusoidally-power-modulated blue laser. The blue laser, which is focused at the center of the drum has a wavelength of 405 nm, an average incident laser power of {0.36~W} and its sinusoidal modulation amplitude is {0.24~W}.
Electrons in the graphene drum are photoexcited, and decay into thermal phonons in less than a picosecond \cite{dawlaty2008measurement,singh2011spectral,vallabhaneni2016reliability} via electron-phonon scattering. Compared to the timescales at which phonons exchange heat (> 0.1 ns), the power transfer from light to lattice vibrations via electron-phonon scattering can thus be considered instantaneous.
 The out-of-plane membrane motion is read out using a photodiode (PD) that detects the reflected intensity of a 633 nm red helium-neon laser with a power of 1.2 mW focused on the center of the membrane, that is modulated by the position-dependent absorption of the graphene membrane \cite{castellanos2013single,dolleman2017amplitude}.  The estimated waist diameter of the focal point is 0.67 $\mu$m for the red laser and 0.57 $\mu$m for the blue laser \cite{dolleman2018transient}. A vector network analyzer (VNA) measures the frequency-dependent amplitude and phase of the signal at the output of a photodetector relative to the modulated blue laser power. The signal is corrected for parasitic phase shifts due to delays in the optical and electronic path using a calibration measurement \cite{PhysRevB.96.165421}, which ensures that the voltage change from the photodiode is linearly proportional to the deflection of the membrane. All experiments are performed at room temperature.  
  
\begin{figure*}[t!]
\centering
\includegraphics{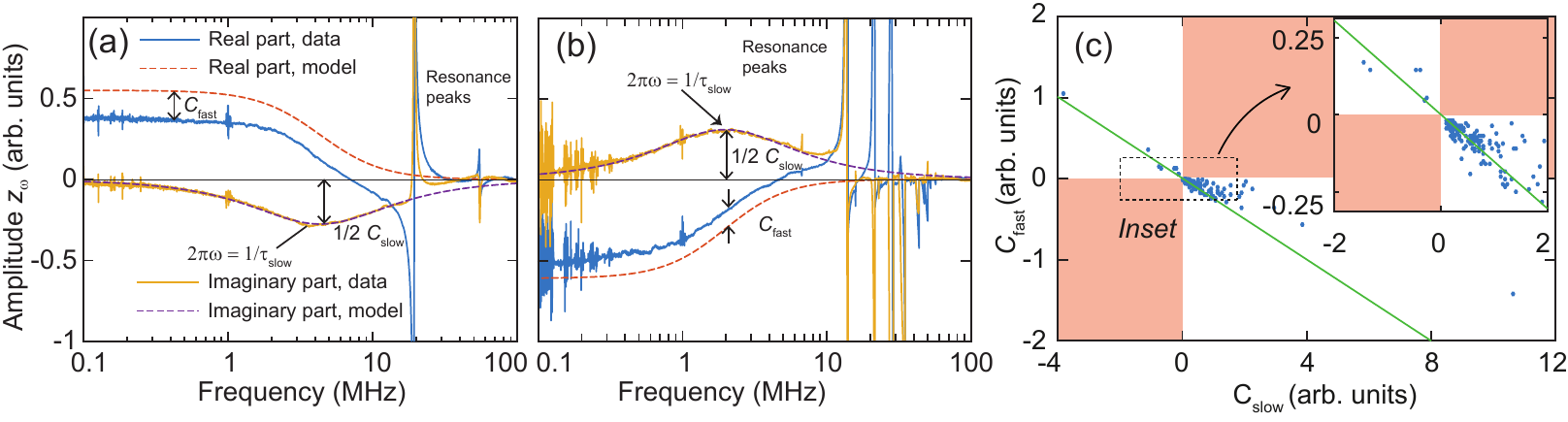}
\caption{Mechanical response of a suspended graphene membrane in response to an intensity-modulated laser. (a) The real and imaginary part of the amplitude of a resonator with a diameter of 4 $\mu$m. (b) The real and imaginary part of a 6-micron diameter drum. (c) Scatter plot with the amplitude of the force $C_{\mathrm{fast}}$  on the vertical axis and the amplitude $C_{\mathrm{slow}}$ on the horizontal axis. Each dot represents a different device with a total of 201 devices measured.  \label{fig:offset}}
\end{figure*}
Figures \ref{fig:offset}(a) and (b) show the real and imaginary amplitude of the membrane's motion as a function of frequency.
To analyse the data, the membrane temperature response $\Delta T$ to a modulated input power $P_{\mathrm{ac}}$ is modeled by the heat equation $\mathcal{C} \Delta \dot{T} +\Delta T/\mathcal{R} = P_{\mathrm{ac}}e^{i \omega t}$, where $\mathcal{C}$ is the effective heat capacitance and $\mathcal{R}$ is the effective thermal resistance of the membrane. $\Delta T$ is the average temperature change over the suspended drum area with respect to the environmental temperature $T_0$, such that the total temperature is given by: $T = T_0 + \Delta T$.  The thermal expansion force is assumed to be proportional to the change in temperature $\Delta T(t)$: $F(t)= \alpha_{\mathrm{eff}} \Delta T(t)$. Furthermore, we assume that far below the mechanical resonance frequency the displacement amplitude $z(t) = F(t)/k$, where $k$ is the effective membrane stiffness. The time-dependent thermal expansion force represented in the frequency domain is then \cite{PhysRevB.96.165421,dolleman2018transient}:  
\begin{equation} 
F_{\omega} e^{i \omega t} \propto z_{\omega} e^{i \omega t}=C_{\mathrm{slow}} \frac{ e^{i\omega t}}{i \omega \tau_{\mathrm{slow}} + 1},
\label{heateq1}
\end{equation}
where $C_{\mathrm{slow}}$ is a constant, representative of the amplitude of the thermal expansion force at low frequencies, used for fitting. $\omega$ is the driving frequency, $\tau_{1} = \mathcal{R}\mathcal{C}$ the thermal time constant and $F_{\omega}$ is obtained by the Fourier transform of $F(t)$. The imaginary part of Eq. \ref{heateq1} has an extremum  with amplitude $C_{\mathrm{slow}}/2$ at radial frequency $\omega = 1/\tau_{\mathrm{slow}}$ as indicated in Figs. \ref{fig:offset}(a) and (b). Only the imaginary part of Eq. \ref{heateq1} is fit to the data, showing good agreement with the experimentally obtained imaginary amplitude. If the real part corresponding to this fit is plotted, however, it is found that below the resonance frequency there is an additional offset $C_{\mathrm{fast}}$, between the real part of Eq. \ref{heateq1} and the measurement, that is almost frequency independent, as indicated in Figs. \ref{fig:offset}(a) and (b).  To quantify the value of $C_{\mathrm{fast}}$ experimentally, the average value of the difference between the real part of the model and the experimental data (see Figs. \ref{fig:offset}(a) and (b)) at frequencies below the resonance frequency is taken. All drums with a negative value of $C_{\mathrm{slow}}$ have a positive offset in the real part $C_{\mathrm{fast}}$ and drums with a positive $C_{\mathrm{slow}}$ have a negative $C_{\mathrm{fast}}$ (Fig. \ref{fig:offset}(c)). We deduce from this correlation between $C_{\mathrm{slow}}$ and $C_{\mathrm{fast}}$ that the offset $C_{\mathrm{fast}}$ is not due to optical cross-talk from the blue laser \cite{PhysRevB.96.165421}, but related to the membrane motion, because optical cross-talk in the setup is independent of the motion of the drum and independent of the sign of $C_{\mathrm{slow}}$. 
\begin{figure}
\includegraphics{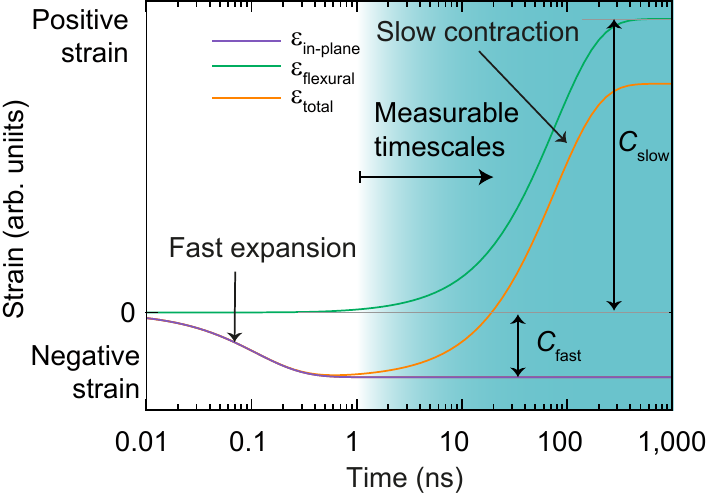}
\caption{ Calculated time-domain representation of the strain if the laser is suddenly switched on at time $t=0$ s.  \label{fig:C1C2}}
\end{figure}

Since the motion that corresponds to the offset $C_{\mathrm{fast}}$ cannot be accounted for by the force in Eq. \ref{heateq1}, it is interpreted as evidence for the existence of an additional second contribution to the thermal expansion force, with a different time-constant $\tau_{\mathrm{fast}}$.
This results in a modified expression for the total thermal expansion force $F_{\omega}$:
\begin{equation} 
F_{\omega} e^{i \omega t} = C_{\mathrm{slow}} \frac{ e^{i\omega t}}{i \omega \tau_{\mathrm{slow}} + 1} + C_{\mathrm{fast}} \frac{ e^{i\omega t}}{i \omega \tau_{\mathrm{fast}} + 1}.
\label{heateq}
\end{equation}
If $\omega \tau_{\mathrm{fast}} \ll 1$, the second contribution to the actuation force produces a constant offset in the real part and does not affect the imaginary part of $F_{\omega}$. Furthermore, a key finding of this work is that $C_{\mathrm{fast}}$ always has an opposite sign with respect to $C_{\mathrm{slow}}$ (Fig. \ref{fig:offset}(c)), meaning that both these forces are counteracting at low frequencies. To show this effect more clearly, the frequency domain response of Eq. \ref{heateq} is converted to a step response in the time domain in Fig. \ref{fig:C1C2}, using typical values of $\tau_{\mathrm{slow}}$ found in this work and an estimate of $\tau_{\mathrm{fast}}$ based on theory \cite{geometricphonon}. Our measurement thus indicates that when a constant heating power is suddenly applied at $t=0$, the membrane initially expands at short timescales $\tau_{\mathrm{fast}}$ and then slowly contracts at longer timescales $\tau_{\mathrm{slow}}$. To explain this observation, the microscopic origin of the thermal expansion contributions in graphene is analyzed in more detail.

The force that actuates the membrane $F(t)$ is directly proportional to the in-plane stress $\sigma(t)$, which is linearly related to the in-plane strain $\epsilon(t)$ by the elasticity matrix. For a membrane clamped around its circumference, this thermally induced strain is related to the internal energies of the phonons and the mechanical properties of the lattice by the equation \cite{ge2016comparative}:\begin{equation}\label{eq:thermalex}
\epsilon(t) = \epsilon_0 - \frac{1}{4  B} \sum\limits_{j}^{} \gamma_j U_j (t),
\end{equation}
where $\epsilon_0$ is the initial pre-strain at the reference temperature when $\Delta T=0$ K, $B$ the bulk modulus, $\gamma_j$ the mode-dependent Grüneisen parameter and $U_j(t)$ the phonon energy per unit volume for phonon mode $i$. Note that $\epsilon(t)$ is the total strain with respect to the initial positive (tensile) pre-strain $\epsilon_0$ at a reference temperature for which $U_j=0$, which is reduced by thermal expansion of the membrane. Thermal expansion of the substrate is neglected in this analysis, since it absorbs less laser power than the graphene and because the volume where the heat can diffuse through is much larger, resulting in negligible temperature changes of the substrate.
 Only the contributions of the acoustic phonon modes are included in the following analysis since the in-plane optical phonon states are not occupied at room temperature and the flexural optical phonons have a Grüneisen parameter close to zero \cite{mann2017negative}. It is well known that the Grüneisen parameter for the flexural phonons $\gamma_{\mathrm{ZA}}$ has a negative sign in graphene, while the Grüneisen parameter for the in-plane longitudinal acoustic ($\gamma_{\mathrm{LA}}$) and transverse acoustic mode ($\gamma_{\mathrm{TA}}$) is positive \cite{ge2016comparative}. 
At low laser modulation frequencies, the internal energy $U_j (t)$ of all phonon modes is in-phase with the blue laser, such that the sign of the thermal expansion force only depends on the sign of the Grüneisen parameter. 
Based on these considerations, the most likely conclusion is that the opposite signs of $C_{\mathrm{slow}}$ and $C_{\mathrm{fast}}$ in the experiments in Fig. \ref{fig:offset} can be attributed to the opposite signs of the in-plane and out-of-plane phonon mode Grüneisen parameters \cite{ge2016comparative}. 
We furthermore hypothesize that the flexural ZA phonons have a longer thermal timescale $\tau_{\mathrm{slow}}$, because they experience a large thermal interface resistance at the edge of the drum \cite{geometricphonon}, while the fast timescale $\tau_{\mathrm{fast}}$ is attributed to the in-plane phonons. The theory that theoretically supports the correctness of this hypothesis is presented in a separate article \cite{geometricphonon}.

The average internal energy $U_j$ of the suspended graphene is modulated by the blue laser with an amplitude that depends on the heat flux absorbed by each mode $P_j$, the mode's Grüneisen parameter $\gamma_j$ and its thermal time constant $\tau_j$. 
We find expressions for the average internal energies $U_j$ in the Supplemental Information \cite{supplemental} and substitute these in Eq. \ref{eq:thermalex} to obtain:
\begin{equation}\label{eq:ratioforces}
\frac{C_{\mathrm{fast}}}{C_{\mathrm{slow}}} = - \frac{(\gamma_{\mathrm{LA}} + \gamma_{\mathrm{TA}})P_{\mathrm{LA+TA}}\tau_{\mathrm{LA+TA}}}{\gamma_{\mathrm{ZA}} P_{\mathrm{ZA}} \tau_{\mathrm{ZA}}} ,
\end{equation}   
where $\tau_{\mathrm{LA+TA}}$ is the fast time constant associated with the in-plane phonons (attributed to $\tau_{\mathrm{fast}}$ for both LA and TA phonons) and $\tau_{\mathrm{ZA}}$ is the slow time constant from the flexural phonons (attributed to $\tau_{\mathrm{slow}}$). Furthermore, an analytical expression for the time constants corresponding to the flexural phonons $\tau_{\mathrm{ZA}}$ is derived by taking only the interaction between the phonon modes at the boundary into account. Using the model in Ref. \cite{geometricphonon}, it is found that the time constant $\tau_{\mathrm{ZA}}$ can be approximated by the expression:
 \begin{equation}\label{eq:tauZA}
 \tau_{\mathrm{ZA}} = \frac{a}{2  \sum \bar{w}_{1z \rightarrow 2r}  c_{\mathrm{ZA}} },
 \end{equation}
where $a$ is the radius of the drum, $\sum \bar{w}_{1z \rightarrow 2r}$ the fraction of ZA phonons that transmit over the boundary towards the environment and $c_{\mathrm{ZA}}$ is the ZA phonon propagation velocity. Both $\sum \bar{w}_{1z \rightarrow 2r}$ and $c_{\mathrm{ZA}}$ are tension-dependent parameters, which increase their value with increasing tension. 

\begin{figure}[t!]
\centering
\includegraphics{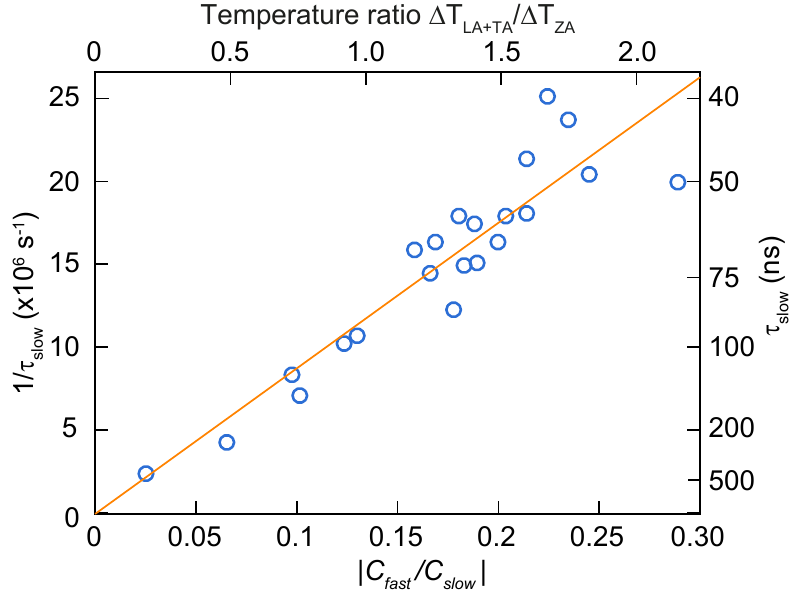}
\caption{Plot of $1/\tau_{\mathrm{slow}}$ versus $|C_{\mathrm{fast}}/C_{\mathrm{slow}}|$ for drums with a diameter of 6 micron, datasets for other diameters can be found in the Supplemental information \cite{supplemental}. The second horizontal axis shows the temperature ratio calculated from Eq. \ref{eq:temprat}.  \label{fig:statistics}}
\end{figure}
 Based on Eq. \ref{eq:ratioforces}, a linear relation between the parameters $-C_{\mathrm{fast}}/C_{\mathrm{slow}}$ and $1/\tau_{\mathrm{ZA}}$ is expected,  assuming that $\tau_{\mathrm{LA+TA}}$ and $P_{\mathrm{LA+TA}}$ are constant because they are relatively unsensitive to tension variations. A study of their correlations is therefore useful as a test for the hypothesis behind Eq. \ref{eq:ratioforces}, since $\tau_{\mathrm{ZA}}$ and $-C_{\mathrm{fast}}/C_{\mathrm{slow}}$ can be extracted independently from the measurement. Figure \ref{fig:statistics}(a) shows a plot with $-C_{\mathrm{fast}}/C_{\mathrm{slow}}$ on the horizontal axis and $1/\tau_{\mathrm{slow}}$ on the vertical axis, for drums with a diameter of 6 $\mu$m. We find a significant linear correlation between $a/\tau_{\mathrm{slow}}$ and $-C_{\mathrm{fast}}/C_{\mathrm{slow}}$, which is in agreement with the models underlying Eqs. \ref{eq:ratioforces}. Interestingly, in the Supplemental information \cite{supplemental}, we find that this correlation is diameter dependent.
 
The ratio $-C_{\mathrm{fast}}/C_{\mathrm{slow}}$ can be used to estimate the degree of thermal nonequilibrium in the system. It is assumed that the changes in modal temperatures of the in-plane phonons are equal $\Delta T_{\mathrm{LA+TA}}  = \Delta T_{\mathrm{LA}} = \Delta T_{\mathrm{TA}}$, based on the results obtained in Ref. \onlinecite{vallabhaneni2016reliability}.
Combining Eq. \ref{eq:ratioforces} with the thermal expansion term $C_j= \gamma_j \rho c_{p,j} \Delta T_j$, where $c_{p,j}$ is the modal specific heat at constant pressure and $\rho$ is the density, we obtain using Eq. \ref{eq:ratioforces}:
\begin{equation}\label{eq:temprat}
\frac{\Delta T_{\mathrm{LA+TA}}}{\Delta T_{\mathrm{ZA}}} = -\frac{C_{\mathrm{fast}}}{C_{\mathrm{slow}}} \frac{\gamma_{\mathrm{ZA}}  c_{p,\mathrm{ZA}}}{\gamma_{\mathrm{LA}}  c_{p,\mathrm{LA}} + \gamma_{\mathrm{TA}} c_{p,\mathrm{TA}}}. 
\end{equation}
The temperature ratio at low frequencies is thus proportional to $-C_{\mathrm{fast}}/C_{\mathrm{slow}}$ with a proportionality constant that can be evaluated from theory. Using $\gamma_{\mathrm{LA}} = 1.06$, $\gamma_{\mathrm{TA}} = 0.40$ and $\gamma_{\mathrm{ZA}} = -4.17$ \cite{mann2017negative}, and the modal specific heats ($c_{p,\mathrm{LA}} = 104$ J/(kg$\cdot$K), $c_{p,\mathrm{TA}} = 225$ J/(kg$\cdot$K), $c_{p,\mathrm{ZA}} = 358$ J/(kg$\cdot$K)) calculated at an environmental temperature of 293.15 K, we obtain: ${\Delta T_{\mathrm{LA+TA}}}/{\Delta T_{\mathrm{ZA}}} = -7.45 C_{\mathrm{fast}}/C_{\mathrm{slow}}$. Using this expression, a histogram of the temperature ratio is constructed as shown in Fig. \ref{fig:statistics}(b).

The average value of the ratio $\Delta T_{\mathrm{LA+TA}}/\Delta T_{\mathrm{ZA}}$ is of the order of 1 (see Fig. \ref{fig:statistics} and the Supplemental information \cite{supplemental}). This is surprising, because the observation that $\tau_{\mathrm{LA+TA}} \ll \tau_{\mathrm{ZA}}$ suggests that the ZA phonons have a very low thermal conductance and therefore, according to Eqs. \ref{eq:ratioforces} and \ref{eq:temprat}, we should expect $\Delta T_{\mathrm{ZA}} \gg \Delta T_{\mathrm{LA+TA}}$. This apparent contradiction between the observed thermal time constants and the temperature ratio is explained the selective electron-phonon coupling in graphene, which causes most of the heat supplied by the laser to end up in the LA and TA phonon bath, while the ZA phonons only receive a small fraction of this heat due to the weak coupling \cite{vallabhaneni2016reliability}. The small value of $\tau_{\mathrm{LA+TA}}/\tau_{\mathrm{ZA}}$ in Eq. \ref{eq:ratioforces} is thus partially compensated by the large value of $P_{\mathrm{LA+TA}}/P_{\mathrm{ZA}}$, thereby causing the temperature of the in-plane and flexural acoustic phonon bath to be in the same order of magnitude. 

In Fig. \ref{fig:statistics}, although a few drums have almost the same value for in-plane and out-of-plane temperature, in many drums large variations in the temperature ratio are observed, with $\Delta T_{\mathrm{LA+TA}}/\Delta T_{\mathrm{ZA}}$ varying from 0.2 to 2.2. This provides evidence for the existence of a strong non-equilibrium thermal state.
According to Eq. \ref{eq:tauZA}, $\tau_{\mathrm{ZA}}$ is tension-dependent, whilst $\tau_{\mathrm{LA+TA}}$ is not expected to be tension dependent. Consequently, according to Eq. \ref{eq:ratioforces}, a linear correlation between $\tau_{\mathrm{slow}}$ and $|C_{\mathrm{fast}}/C_{\mathrm{slow}}|$ as found in Fig. \ref{fig:statistics}(a) shows that the large variations in the temperature ratio are dominated by device-to-device variations in the pre-tension via its effect on the thermal time constant $\tau_{\mathrm{ZA}}$. Similar large variations in $\tau_{\mathrm{ZA}}$ have been observed in our previous work \cite{PhysRevB.96.165421}. Some devices deviate from this linear correlation (see Supplemental Information \cite{supplemental}),  this might suggest that other effects such as wrinkles and other imperfections are also playing a role in the variations in the temperature ratio. 

The observed nonequilibrium effect has important consequences for the interpretation of thermal measurements on graphene, as it becomes difficult to determine the contribution of each phonon mode to the thermal conductivity. Moreover, the extracted thermal conductivity obtained from the classical heat equation can become geometry dependent. This affects any suspended graphene device that is locally heated, due to the inherent selective electron-phonon coupling and weak interaction between the phonon modes.

 Other two-dimensional materials are expected to show similar effects as observed in this work if they exhibit weak mode interaction and a large negative Grüneisen parameter for the flexural acoustic phonons. This might hold for other monatomic two-dimensional materials at room temperature \cite{ge2016comparative}, but also transition metal dichalcogenides such as MoS$_2$, MoSe$_2$ and WS$_2$ at low temperatures ($< 100$ K) \cite{peng2016thermal}. 

To conclude, we have presented evidence that the motion of opto-thermally excited graphene resonators is the result of two counteracting contributions in the thermal expansion force. The amplitude of these contributions provides information on the ratio of the effective temperatures of the thermal baths of the in-plane and flexural acoustic phonons and based on a model it is shown that they are at different temperatures.  These thermal nonequilibrium effects should be considered in the interpretation of the thermal conductivity measurements of 2D materials. Moreover, they are shown to lead to an unconventional time-dependent sign of thermal expansion forces in graphene.

\begin{acknowledgements}
The authors thank Applied Nanolayers B.V. for supply and transfer of the single-layer graphene. We furthermore thank C. Stampfer, J. Sonntag and J.E. Sader for fruitful discussions. This work is part of the research programme Integrated Graphene Pressure Sensors (IGPS) with project number 13307 which is financed by the Netherlands Organisation for Scientific Research (NWO).
The research leading to these results also received funding from the European Union's Horizon 2020 research and innovation programme under grant agreement No 785219 Graphene Flagship. 
\end{acknowledgements}

\newpage
~
\newpage

\begin{widetext}
\section*{Supplemental Information}
\section*{S1: Diameter dependence}
\begin{figure}[h!]
\centering
\includegraphics[width = \linewidth]{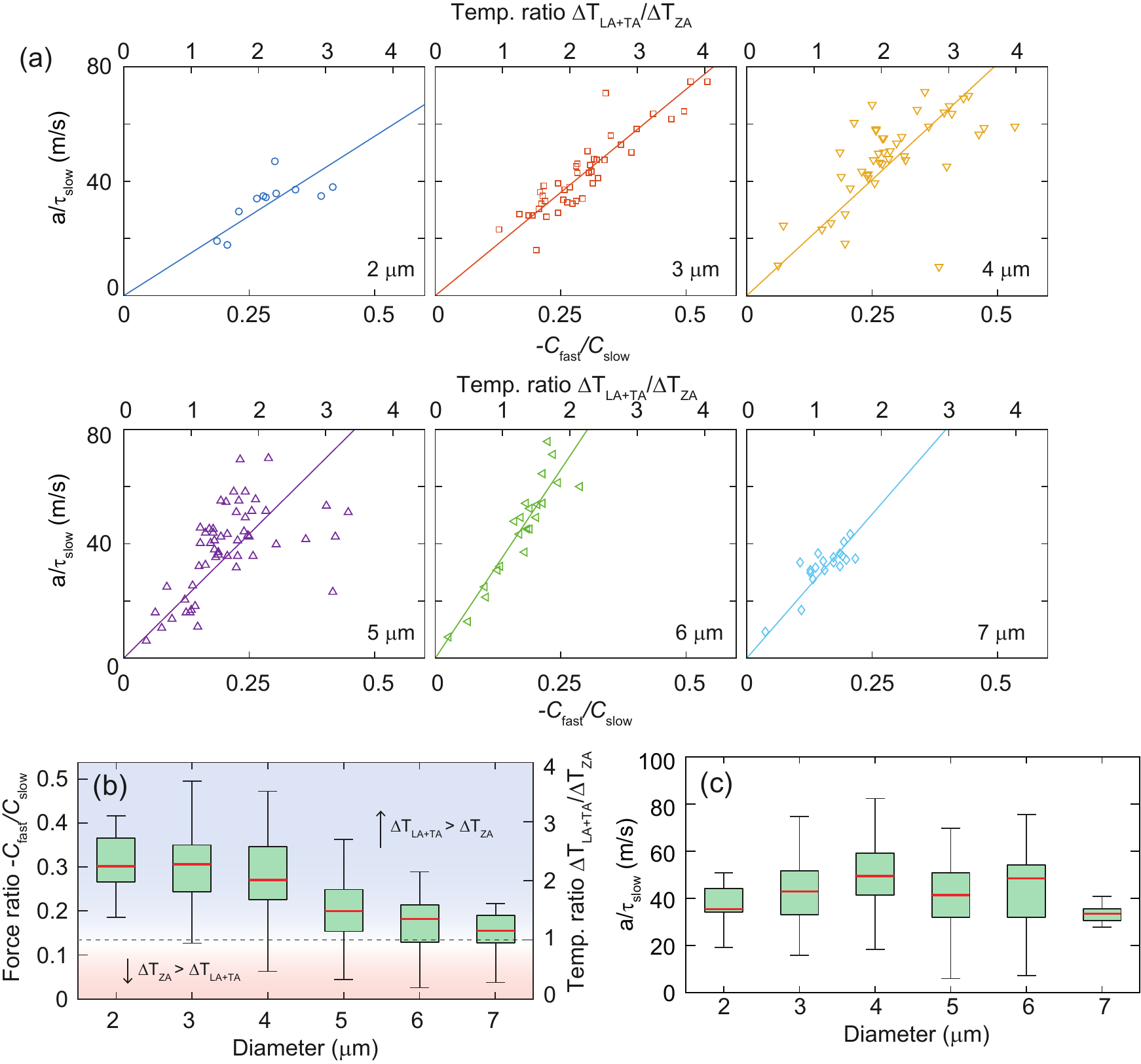}
\caption{(a) Scatter plot of $a/\tau_{\mathrm{slow}}$ versus $|C_{\mathrm{fast}}/C_{\mathrm{slow}}|$ for each diameter. A straight line presents a least square fit to the data, and reveals a diameter dependent slope. (b) Box plot of the ratio between the thermal expansion forces $-C_{\mathrm{fast}}/C_{\mathrm{slow}}$, the second vertical axis shows the calculated temperature ratio ${\Delta T_{\mathrm{LA+TA}}}/{\Delta T_{\mathrm{ZA}}}$. A horizontal dashed line shows where ${\Delta T_{\mathrm{LA+TA}}}={\Delta T_{\mathrm{ZA}}}$, drums that fall below this line show a higher temperature of the ZA phonons while drums above this line show a higher temperature of the LA and TA phonons.  (c) Boxplot of $a/\tau_{\mathrm{slow}}$ as a function of drum diameter.  \label{fig:statistics}}
\end{figure}

Figure \ref{fig:statistics}(a) shows the correlation between the parameters $a/\tau_{\mathrm{slow}}$ and $-C_{\mathrm{fast}}/C_{\mathrm{slow}}$. All diameters show a significant linear correlation between these parameters. However, the slopes of the straight lines that are fit to the data are diameter dependent. To investigate this diameter dependence further, a boxplot of $-C_{\mathrm{fast}}/C_{\mathrm{slow}}$ and ${\Delta T_{\mathrm{LA+TA}}}/{\Delta T_{\mathrm{ZA}}}$ for each diameter is made as shown in Fig. \ref{fig:statistics}(b), and a box plot of $a/\tau_{\mathrm{slow}}$ is made.   Comparing the diameter dependence of $-C_{\mathrm{fast}}/C_{\mathrm{slow}}$ to the diameter dependence of $a/\tau_{\mathrm{slow}}$, the largest relative change in the mean values is observed in $-C_{\mathrm{fast}}/C_{\mathrm{slow}}$, suggesting this is the underlying mechanism behind the diameter dependent slope in Fig. \ref{fig:statistics}(a).

Our model in Ref. \cite{geometricphonon} suggests no diameter-dependence of $a/\tau_{\mathrm{slow}}$ should occur, if the tension in the drums is not diameter dependent. No significant diameter-dependence of $a/\tau_{\mathrm{slow}}$ is discovered in Fig. \ref{fig:statistics}(b), except for the 7-micron diameter drums where small values of $\tau_{\mathrm{slow}}$ appear to be missing. Drums with a small value of $\tau_{\mathrm{slow}}$ are expected to have a large tension. Therefore, this might be an indication of a selective breaking mechanism for these drums, where drums with large tensions are more likely to fail. Apart from the 7-micron drums, each diameter shows a considerable spread in $a/\tau_{\mathrm{slow}}$, which may be attributed to device-to-device variations in the tension of the drums, that alter $\tau_{\mathrm{slow}}$. This is also the underlying reason for spread along the straight lines in Fig. \ref{fig:statistics}(a). 

The diameter dependence of $-C_{\mathrm{fast}}/C_{\mathrm{slow}}$ is unexpected, as our model \cite{geometricphonon} predicts that $|C_{\mathrm{fast}}/C_{\mathrm{slow}}|$ increases as a function of diameter, while our data in Fig. \ref{fig:statistics}(b) shows that it decreases. Several scenarios are investigated in Ref. \cite{geometricphonon} in order to explain this effect. The most probable explanation is that the diameter dependence is caused by ballistic effects in phonon transport. 

\section*{S2: Derivation of the expression for the ratio $C_{\mathrm{fast}}/C_{\mathrm{slow}}$}
Here the expression for the ratio between the two contributions to the thermal expansion force $C_{\mathrm{fast}}/C_{\mathrm{slow}}$ is derived. We assume for each force that $\omega \ll \tau_i$. From eq. (3) in the main section of the paper we have:
\begin{equation}\label{eq:thermalex}
\epsilon = \epsilon_0 - \frac{1}{4 B} \sum\limits_{i}^{} \gamma_j U_j ,
\end{equation}
Linearizing the system for small $\Delta T$:
\begin{equation}\label{eq:UdeltaT}
U_j = \rho c_{p,j} \Delta T_j.
\end{equation}
Since the thermal expansion forces $C_{\mathrm{slow}}$ and $C_{\mathrm{fast}}$ are proportional to the respective phonon bath contributions to the thermal strain $\epsilon$, it follows from Eqs. \ref{eq:thermalex} and \ref{eq:UdeltaT} that: 
\begin{equation}
\frac{C_{\mathrm{fast}}}{C_{\mathrm{slow}}} = \frac{\gamma_{\mathrm{LA}} \rho c_{p,\mathrm{LA}} \Delta T_{\mathrm{LA}} + \gamma_{\mathrm{TA}} \rho c_{p,\mathrm{TA}} \Delta T_{\mathrm{TA}}}{\gamma_{\mathrm{ZA}} \rho c_{p,\mathrm{ZA}} \Delta T_{\mathrm{ZA}} }
\end{equation}
Since $\Delta T_j = R_{B,j} P_{ac,j}/A$, where $A$ is the area of the circumference and $R_{B,j}$ the thermal interface resistance, we can write:
\begin{equation}\label{eq:ratio1}
\frac{C_{\mathrm{fast}}}{C_{\mathrm{slow}}} = \frac{\gamma_{\mathrm{LA}} \rho c_{p,\mathrm{LA}} R_{B,\mathrm{LA}} P_{ac,\mathrm{LA}} + \gamma_{\mathrm{TA}} \rho c_{p,\mathrm{TA}}  R_{B,\mathrm{TA}} P_{ac,\mathrm{TA}}}{\gamma_{\mathrm{ZA}} \rho c_{p,\mathrm{ZA}}  R_{B,\mathrm{ZA}} P_{ac,\mathrm{ZA}} },
\end{equation}
Now it is convenient to convert the specific heat $\rho c_{p,j}$ into the modal heat capacity $\mathcal{C}_j$, from \cite{PhysRevB.96.165421}:
\begin{equation}
\mathcal{C}_j = \rho c_{p,j} h_g \pi a^2,
\end{equation}
For the thermal resistance:
\begin{equation}
\mathcal{R}_j = \frac{R_B}{h_g \pi a^2}.
\end{equation}
Using this and the relation $\tau_j = \mathcal{R}_j\mathcal{C}_j$, we arrive at:
\begin{equation}\label{eq:ratio2}
\frac{C_{\mathrm{fast}}}{C_{\mathrm{slow}}} = \frac{\gamma_{\mathrm{LA}} \tau_{\mathrm{LA}} P_{ac,\mathrm{LA}} + \gamma_{\mathrm{TA}} \tau_{\mathrm{TA}} P_{ac,\mathrm{TA}}}{\gamma_{\mathrm{ZA}}\tau_{\mathrm{ZA}} P_{ac,\mathrm{ZA}} },
\end{equation}
Since it is assumed the in-plane LA and TA phonons are at the same temperature, we can write this expression as:
\begin{equation}\label{eq:ratioforces}
\frac{C_{\mathrm{fast}}}{C_{\mathrm{slow}}} = \frac{(\gamma_{\mathrm{LA}} + \gamma_{\mathrm{TA}})P_{\mathrm{LA+TA}}\tau_{\mathrm{LA+TA}}}{\gamma_{\mathrm{ZA}} P_{\mathrm{ZA}} \tau_{\mathrm{ZA}}}.
\end{equation}   
\end{widetext}

\end{document}